\documentclass[pra,showpacs,twocolumn,superscriptaddress,longbibliography]{revtex4-2}
\usepackage{amsmath}
\usepackage{amssymb}
\usepackage{graphicx}
\usepackage{dcolumn}
\usepackage{bm}
\usepackage{amsfonts}
\usepackage{subfigure}
\usepackage{float}
\usepackage{color}
\usepackage{xcolor}
\usepackage{braket}
\usepackage{hyperref}
\usepackage{comment}

\makeatletter

\renewcommand{\figurename}{\textbf{Fig.}}
\renewcommand{\fnum@figure}
{\textbf{\figurename~\thefigure}}

\def\maketitle{
\@author@finish
\title@column\titleblock@produce
\suppressfloats[t]}
\makeatother
\usepackage[normalem]{ulem}

\hypersetup{pdfborder = {0 0 0}}

\begin{document}
\title{Direct observation of long-range many-body coherence in quasi-one-dimensional attractive Bose gases}

\author{Hikaru Tamura}
\email{tamurah@ims.ac.jp}
\altaffiliation{Present address: Institute for Molecular Science, Okazaki, Aichi 444-8585, Japan}
\affiliation{Department of Physics and Astronomy, Purdue University, West Lafayette, IN 47907, USA}

\author{Sambit Banerjee}
\affiliation{Department of Physics and Astronomy, Purdue University, West Lafayette, IN 47907, USA}

\author{Rongjie Li}
\affiliation{Department of Physics and Astronomy, Purdue University, West Lafayette, IN 47907, USA}

\author{Panayotis Kevrekidis}
\affiliation{Department of Mathematics and Statistics, University of Massachusetts Amherst, Amherst, Massachusetts 01003-9305, USA}
\affiliation{Department of Physics, University of Massachusetts Amherst, Amherst, Massachusetts 01003, USA}

\author{Simeon I. Mistakidis}
\affiliation{Department of Physics and LAMOR, Missouri University of Science and Technology, Rolla, MO 65409, USA}

\author{Chen-Lung Hung}
\email{clhung@purdue.edu}
\affiliation{Department of Physics and Astronomy, Purdue University, West Lafayette, IN 47907, USA}
\affiliation{Purdue Quantum Science and Engineering Institute, Purdue University, West Lafayette, IN 47907, USA}

\date{\today }

\begin{abstract} 
Macroscopic coherence is an important feature of quantum many-body systems exhibiting collective behaviors, with examples ranging from atomic Bose-Einstein condensates, and quantum liquids to superconductors. Probing many-body coherence in a dynamically unstable regime, however, presents an intriguing and outstanding challenge in out-of-equilibrium quantum many-body physics. Here, we experimentally study the first- and second-order coherence of degenerate quasi-one-dimensional (1D) Bose gases quenched from repulsive to modulationally unstable attractive interaction regimes. The resulting dynamics, monitored by in-situ density and matter-wave interference imaging, reveals phase-coherent density wave evolutions, clearly distinguished from iconic soliton trains previously observed in attractive gases. This arises from the interplay between noise-amplified density modulations and dispersive shock waves of broad interest in nonlinear physics, plasmas, granular systems, and beyond. At longer times, the gases become phase-scrambled, exhibiting a finite correlation length. Interestingly, following an interaction quench back to the repulsive regime, we observe that quasi-long-range coherence can be spontaneously re-established. This captivating rephasing dynamics can be attributed to the nucleation and annihilation of density defects in the quasi-1D geometry. These results shed light on out-of-equilibrium phase coherence in quantum many-body systems in a regime where beyond mean-field effects may arise and theoretical approaches have not been well-established.
\end{abstract}

\maketitle

Long-range phase coherence is a defining feature of Bose-Einstein condensates (BECs) \cite{anderson1995observation} and superfluidity \cite{raman1999evidence}. It plays a critical role not only in equilibrium but also in out-of-equilibrium conditions dictating quantum superposition and interference since it quantifies crosstalk of atoms at long distances. A global quench in ultracold gases offers an ideal setting for studying the interplay between long-range phase coherence and collective or nonlinear excitations. Such quenches have enabled observations of prethermalization~\cite{gring2012relaxation,langen2013local} and recurrent long-range coherence~\cite{rauer2018recurrences} in one-dimensional (1D) Bose gases, phase scrambling and rephasing with collective modes in a dipolar supersolid~\cite{ilzhofer2021phase}, dynamics of vortex-antivortex pairs across the BKT transition~\cite{sunami2023universal}, phase-coherent domain formation in Bose gases quenched across the BEC transition~\cite{corman2014quench,navon2015critical} and across the Mott-insulator-to-superfluid phase  transition~\cite{chen2011quantum} in connection with the Kibble–Zurek mechanism~\cite{kibble1976topology,kibble1980some}. Uncovering the quantum dynamics requires direct, spatially resolved access to phase information for which, matter-wave interferometry~\cite{hadzibabic2006berezinskii} has been used as a key experimental tool, revealing the presence of out-of-equilibrium phase correlations~\cite{gring2012relaxation,langen2013local,rauer2018recurrences,corman2014quench,sunami2023universal,navon2015critical}. Although important aspects of local phase coherence have been explored in repulsively interacting gases, its direct experimental observation in the attractive interaction regime remains elusive. 

Attractive gases and their phase coherence, unlike their stable repulsive counterparts, are known to suffer modulational instability (MI). This mechanism triggers the amplification of small fluctuations and leads to the formation of localized solitonic structures~\cite{Strecker2002,Strecker2003,nguyen2017formation,robbins}. In $D>1$ dimensions, growing waves in MI introduce wave collapse \cite{gerton2000direct,donley2001dynamics,eigen2016observation,chen2020observation} at a dominant unstable length scale determined by the interaction strength, i.e., the healing length. The wave collapse is responsible for spontaneously forming bright solitons and arrays thereof \cite{cornish2006formation,nguyen2017formation,chen2020observation}. Bright solitons have by now been showcased to be ubiquitously possible to produce across harmonically trapped atomic gases. These include $^7$Li~\cite{kasevich1}, where solitonic pairs were realized, $^{85}$Rb~\cite{Marchant2013}, for examining soliton-barrier interactions, and a solitonic atom interferometer~\cite{robbins1}, as well as Cs~\cite{mevznarvsivc2019cesium, di2019excitation} and $^{39}$K~\cite{aspect1,sanz2022interaction}.

\begin{figure*}[t!]
\centering
\includegraphics[width=0.9\textwidth]{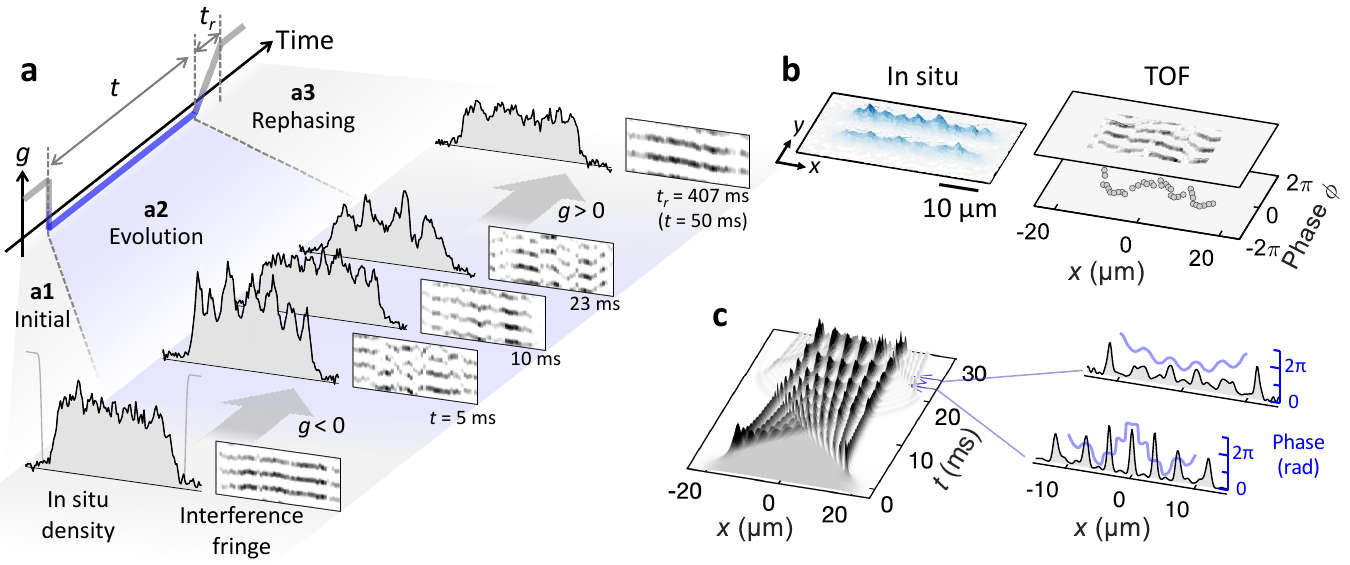}
\caption{\textbf{Dynamical density modulations and phase coherence of quasi-1D attractive Bose gases in a box.} {\textbf{a}}, A repulsively interacting atomic superfluid (a1) is quenched to the weakly attractive regime where it is held for time $t$ (a2). The interaction is then slowly brought back to its initial repulsive value, triggering dynamical rephasing (a3). {\textbf{b}}, In each experimental realization, two quasi-1D gases are prepared in parallel. They are imaged in situ or released to produce matter-wave interference via an in-plane time-of-flight (TOF) before imaging. A local relative phase $\phi$ is obtained from the extracted interference fringes. Prototypical in-situ densities and interference patterns right before the quench (a1), for different hold times (a2), and after a long ramp (a3) are displayed in \textbf{a}. {\textbf{c}}, 1D GPE simulation of an ideal wavefunction confined in a Gaussian wall box and quenched from repulsive to attractive interactions. The black (blue) solid lines represent density (phase) profiles at the times indicated by the arrows.}
\label{fig:fig1}
\end{figure*}

Beyond spatially separated solitons, attractive gases in 1D could exhibit very rich out-of-equilibrium behavior during the nonlinear stage of MI \cite{zakharov2013nonlinear,GinoDion,mossman2024nonlinear}. It is known that 1D Schr\"odinger models with self-focusing cubic-nonlinearity feature infinitely many conservation laws. Indeed, the associated integrability enables the identification of numerous families of special solutions, including those of spatially or temporally periodic breather solutions~\cite{akhmediev1986modulation,peregrine1983water}, whose formation process~\cite{akhmediev2009extreme,el2016dam} has recently attracted great attention~\cite{soto2016integrable,akhmediev2016breather}. Their evolution represents dynamical energy exchange between unstable modes within a modulationally unstable background. This manifests itself in the phase sector as transitions between complete $\pi$ phase jumps and a homogeneous phase profile. As a direct consequence of integrability in 1D, the conservation laws prevent breathers from decaying into phase uncorrelated solitons \cite{mahnke2012possibility}. Phase coherence thus provides a decisive factor to rule out the existence of phase-uncorrelated excitations. However, out-of-equilibrium phase coherence beyond solitary waves in attractive many-body systems, especially in quantum gas experiments, remains largely unexplored.

Here, we report the observation of dynamical evolution of many-body coherence in quasi-1D Bose gases quenched from repulsive to attractive interactions. The emergent dynamics of our box-trapped quasi-1D gases differs significantly from harmonically trapped ones; see also Supplementary Material (SM) for relevant comparisons. As illustrated in Fig.~\ref{fig:fig1}a, the gas is confined by an elongated optical box trap, featuring prethermalized density fluctuations over a homogeneous background (see~SM) and sharp density gradients at the edges in the initial superfluid phase (a1). After an interaction quench (a2), the gas evolves towards displaying not only noise-amplified density modulations in the bulk but also shock waves arising from the edges. By monitoring local density and phase fluctuations through in-situ imaging and matter-wave interference, respectively, we observe the resulting dynamical generation of density modulations with preserved phase coherence. Interestingly, this emerging signature represents neither solitons nor breathers but non-insulated density and phase profiles as remnants of the nonlinear stage of MI in quasi-1D, as visualized by the first- and second-order coherence. At longer hold times, we find that the phase correlation length scale is shortened to the healing length, which we attribute to global dephasing. 

\begin{figure*}[t!]
\centering
\includegraphics[width=0.9\textwidth]{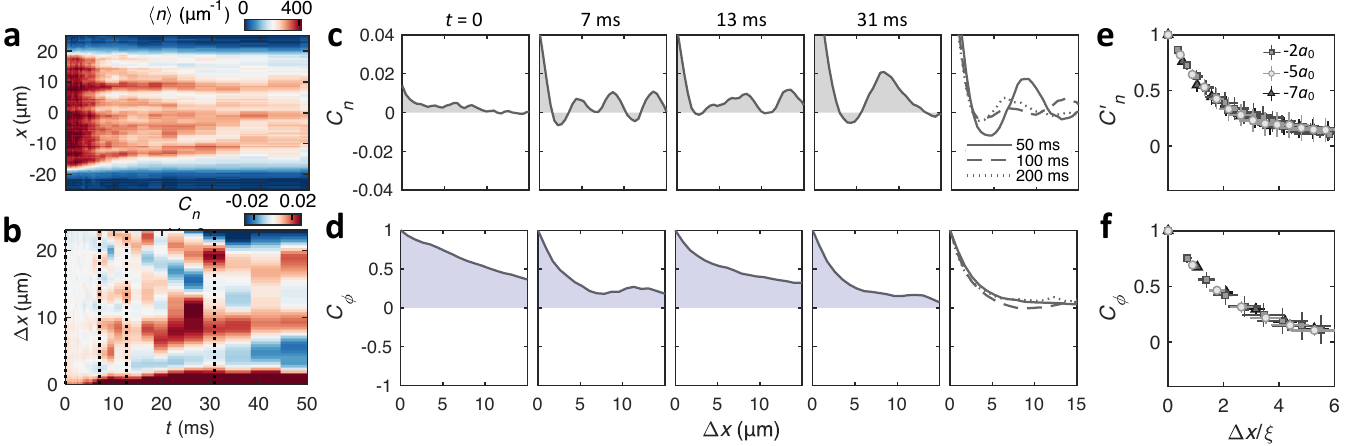}
\caption{\textbf{Observation of dynamical density modulations and phase coherence of quenched quasi-1D Bose gases.} {\textbf{a}}, Time evolution of the sample-averaged density profile after a quench of s-wave scattering length from $a\approx 105a_0$ to $-5a_0$. The density is integrated over the transverse directions and plotted along the $x$-axis. {\textbf{b}}, Dynamics of density-density correlation function $C_{n} (\Delta x)$ with two-point separation $\Delta x$. {\textbf{c}}, $C_{n} (\Delta x)$ at the characteristic times marked by dashed lines in {\textbf{b}}, and at the indicated long hold times. {\textbf{d}}, Two-point phase correlation function $C_{\phi} (\Delta x)$ at the same times. {\textbf{e}}, Normalized density-density correlation function $C’_{n}=C_{n}(\Delta x)/C_{n}(0)$ and {\textbf{f}}, $C_{\phi}$ versus rescaled length $\Delta x / \xi$ at three different post-quench scattering lengths. $C'_n$ and $C_{\phi}$ are time-averaged over samples measured after hold time $t=50\,\rm{ms}$. The vertical error bars show the standard deviation from the time-averaged values. The horizontal error bars stem from the uncertainty in the determination of the scattering length.}
\label{fig:fig2}
\end{figure*}

Exploiting a reverse quench, that is, ramping the interaction from attractive back to its initial repulsive value, the system returns to a modulationally stable regime (a3). We observe that quasi-long-range phase coherence re-establishes progressively at slower ramps. Both our experiments and mean-field simulations indicate that density modulations at the attractive interaction convert to dark solitary waves in the repulsive interaction regime. Our studies on defect dynamics suggest the importance of extended dimensionality beyond pure 1D phenomenology, where we attest that mechanisms for defect annihilation~\cite{Becker_2013,kwon2014relaxation,kanai2021true} are responsible for reinstating the global phase coherence.

Our experiment begins with loading a nearly pure cesium BEC into two parallel quasi-1D boxes; see SM for details. Each box has length $l=40\,\rm{\mu m}$ along the longitudinal ($x$-)axis. The transverse trap frequencies are $(\omega_{y}, \omega_{z})\approx 2\pi\times (68, 2.2\times 10^{3})\,\rm{Hz}$, giving harmonic oscillator lengths $(l_{y}, l_z)\approx (1.1, 0.18)\,\rm{\mu m}$ and high trap aspect ratios $(\frac{l}{l_y},\frac{l}{l_z})\approx (36, 220)$. After an initial trap loading at a $s$-wave scattering length of $105a_{0}$, where $a_0$ is the Bohr radius, we quench to a weakly negative value $a\approx -5a_{0}$ by controlling a magnetic Feshbach resonance~\cite{chin2010feshbach}. The post-quench interaction energy $n|g|\approx h\times 76\,\rm{Hz}$ is comparable to the trap vibrational energy $\hbar\omega_{y}$, leading to a transverse motion that damps out in the first 3~ms followed by a gradual atom loss (SM). Here, $g\approx 2\hbar^{2} a/(ml_{y}l_{z})$ is the 1D interaction parameter, $n \approx 400\,\rm{\mu m}^{-1}$ is the effective 1D density, $h=2\pi \hbar$ is the Planck constant, and $m$ is the atomic mass. Because of a tighter confinement, the atomic motion along the $z$-axis remains approximately in the ground state throughout the quench dynamics. 

The density profiles of the quenched gases are monitored by in-situ density imaging (left panel in Fig.~\ref{fig:fig1}b). To probe phase coherence, we adapt a matter-wave-interference technique \cite{langen2015experimental, schweigler2017experimental}. After a hold time $t$, we perform time-of-flight (TOF) imaging in the $x$-$y$ plane. The parallel gases expand quickly along the $y$-axis due to large aspect ratio and interfere, yielding information about their relative phase $\phi (x)$ longitudinally; see Fig.~\ref{fig:fig1}b.

Quench dynamics of a 1D homogeneous quantum gas with sharp edges can be mapped to a recently studied dam-break problem~\cite{el2016dam, sharan2025breaking}. Upon an interaction quench the initial sharp density gradients at the edges produce a pair of dispersive shock waves (DSWs)~\cite{EL201611} that evolve through the nonlinear stage of MI~\cite{GinoDion}. The DSW is a rapidly oscillating wave structure that is self-similarly modulated at a slower scale and is sustained by the competition of wave steepening and dispersion. More broadly, this setting is closely related to the so-called Riemann problems involving initial density or velocity jumps that have recently attracted extensive interest in dispersive nonlinear systems~\cite{biondini2018riemann}. 

To understand density and phase evolutions purely originating from DSWs, we first deploy numerical simulations of the 1D mean-field Gross-Pitaevskii equation (GPE). We start with the ground state of the gas confined in a box with Gaussian-wall width $3~\rm{\mu m}$ approximating the experimental potential, and the scattering length is quenched from $105a_{0}$ to $-4.6a_{0}$. The produced DSWs interfere and form dynamical density patterns (left panel in Fig.~\ref{fig:fig1}c) with a changing spatial periodicity. This temporal evolution presents an alternating sequence between quasi-periodic density peaks with large phase modulations (bottom) and less undulated density and phase profiles (top in the right panel). In our computations, each resulting peak near the center represents a phase-correlated Peregrine soliton-like structure~\cite{peregrine1983water}. We confirm the presence of a pair of near-$\pi$ phase jumps across its core and tails; see Fig.~\ref{fig:fig1}c. Similar structures have been theoretically predicted in the integrable cubic focusing  GPE model~\cite{el2016dam}. A Peregrine soliton was recently observed in a two-component repulsive gas~\cite{romero2024experimental}. 

In our experiment, initial density fluctuations also seed MI, leading to the interplay between noise-amplified modulations and the DSWs. The former is randomly seeded in each realization, washing out the aforementioned localized structures in the sample-averaged density (Fig.~\ref{fig:fig2}a). The density profile near the boundaries, however, tends to shrink inwards shortly after the quench, indicating the emergence of the DSWs.

To analyze the quench dynamics reflecting both the DSWs and the noise-induced density modulations, we measure the density-density correlation function
\begin{equation}
C_{n}(\Delta x) = \frac{\Braket{n(x) n(x')}}{\Braket{n(x)}\Braket{n(x')}}-1,
\end{equation}
where $\Delta x=|x-x'|$ and $\braket{\dots}$ denotes ensemble (around 20 samples) as well as spatial averaging. The correlation function estimates the second-order coherence, yielding density structure information--bunching ($C_{n}>0$) or antibunching ($C_{n}<0$) of densities separated by $\Delta x$. Figure~\ref{fig:fig2}b shows the observed dynamics of $C_{n}$ within a spatial window of $|x|, \,|x'|\leq 15\,\rm{\mu m}$, representing density waves with dynamically changing periodicity for $t\lesssim 30$~ms. At early times $t\sim 7\,\rm{ms}$, density modulations with a short periodicity of $\sim 4\,\rm{\mu m}$ emerge, being clearly visible in the correlation profile depicted in Fig.~\ref{fig:fig2}c. This formation is attributed to fluctuation-seeded MI whose length scale $2\pi \xi \approx 4.4\,\rm{\mu m}$ is consistent with the observed periodicity, where $\xi = \hbar/\sqrt{2mn|g|}$ is the healing length~\cite{salasnich2003modulational, nguyen2017formation, chen2020observation}. At $t\sim 13\,\rm{ms}$ roughly corresponding to the time scale of the DSW arrival at the box center (SM), the noise-amplified modulation interacts with the DSWs. Their competition is evidenced through the manifestation of density modulations involving multiple spatial frequencies, which indeed appears as a common phenomenon as predicted by the 1D self-focusing nonlinear Schr\"odinger equation~\cite{el2016dam}. At $t\gtrsim31\,\rm{ms}$, a density modulation of $\sim$10~$\mu$m periodicity prevails. We find that, at later times, the density-density correlation continues to evolve (see the last column in Fig.~\ref{fig:fig2}c).

To further visualize the many-body phase coherence characterizing the formation of density modulations, we evaluate the two-point phase correlation function
\begin{equation}
C_{\phi}(\Delta x) = \left\langle {\mathrm{Re}} \left[  e^{i\phi(x) - i \phi(x+\Delta x)} \right] \right\rangle \,,
\end{equation}
which reveals the first-order coherence of each gas when they are occupied primarily with quasi-condensate atoms (SM). At $\Delta x > \xi$, non-zero $C_\phi(\Delta x)$ indicates finite-range phase correlations; $C_\phi \approx 0$ when the gas becomes phase uncorrelated or filled with random phase jumps. As shown in Fig.~\ref{fig:fig2}d, the measurement at $t=0$ displays quasi-long-range coherence, decaying at a length $L\approx16~\mu$m that is consistent with interfering two prethermalized 1D gases after splitting from a condensate \cite{gring2012relaxation,langen2013local}; see also~SM. After the interaction quench $t>0$, we observe remarkable dynamical coherence patterns. First, the gas retains finite-range phase correlations even after a periodic density modulation has emerged ($t\gtrsim7~\rm{ms}$). Second, when multiple spatial frequencies compete in the density modulation ($t\sim 13~\rm{ms}$), we observe that the range of phase coherence transiently increases. At $t\gtrsim 31\,\rm{ms}$, the phase correlation gradually decreases to a shorter length scale. After sufficiently long hold times, the length scales of both density and phase correlations slowly approach the healing length $\xi$, as shown in Figs.~\ref{fig:fig2}e and f, respectively. We interpret this as the gas is filled with seemingly random density modulations and phase jumps, making the averaged correlation length shrink to $\sim \xi$.  

\begin{figure}[t]
\centering
\includegraphics[width=0.45\textwidth]{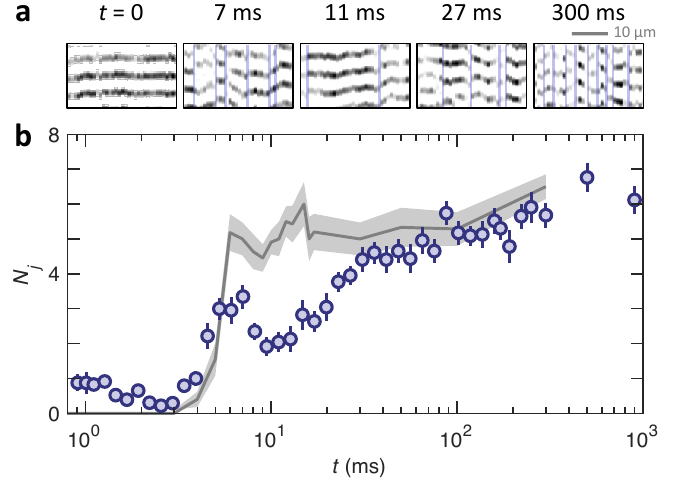}
\caption{
\textbf{Observation of phase slips.} {\textbf{a}}, Interference fringes after the quench to the attractive scattering length $-5a_{0}$ discussed in Fig.~\ref{fig:fig2}. The vertical lines mark detected phase slips. {\textbf{b}}, Number of phase jumps $N_{j}$ versus the hold time $t$ after the interaction quench. Error bars indicate the standard error of the mean. Solid line shows the simulated $N_{j}$ using the 2D GPE ($\geq 6$ samples for a given $t$). The shaded area represents the standard error of the mean.}
\label{fig:phase_slip}
\end{figure}

\begin{figure}[t!]
\centering
\includegraphics[width=0.45\textwidth]{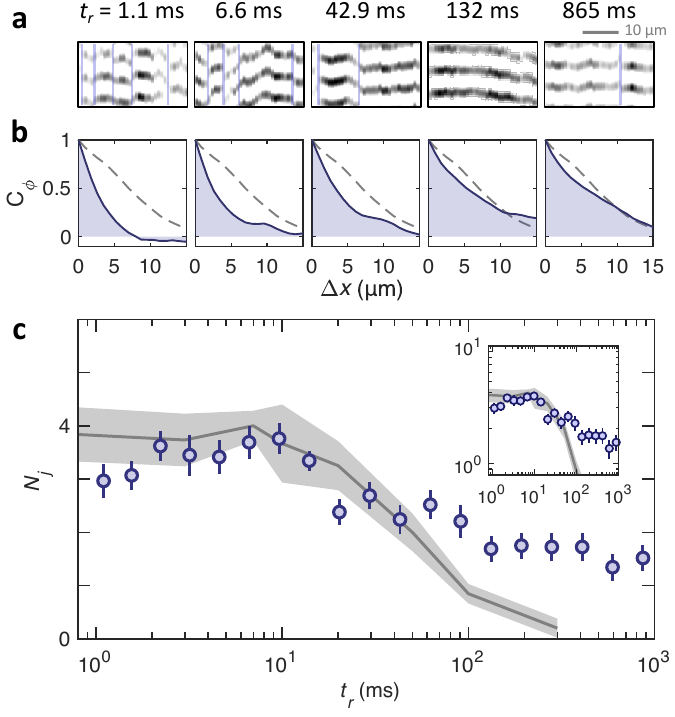}
\caption{
\textbf{Observation of rephasing dynamics.} {\textbf{a}}, Interference fringes after holding the scattering length at $-5a_{0}$ for $50~\rm{ms}$ followed by quench-up to the initial repulsion $105a_{0}$ within the ramp time $t_{r}$. The vertical lines mark the locations of the detected phase slips. {\textbf{b}}, The corresponding two-point phase correlation function $C_{\phi}$. The dashed line in each panel represents the initial $C_{\phi}$ before quenching to attractive interactions. {\textbf{c}}, Number of phase jumps $N_{j}$ as a function of $t_{r}$. Errorbars denote the standard error of the mean. Solid line represents $N_{j}$ identified in 2D GPE simulations ($\geq 8$ samples at each $t_{r}$). The shaded area indicates the standard error of the mean of the simulations. Inset shows the same data along with the corresponding 2D GPE simulations in log-log scale.
}
\label{fig:fig4}
\end{figure}

Throughout the evolution, we expect phase jumps to accompany the formation of density modulations, as illustrated in Fig.~\ref{fig:fig1}c. The evolution and loss of long-range phase coherence are traced back to the development of phase jumps. To visualize this connection, we count the number of fast phase slips, $N_j$, in the unwrapped $\phi(x)$ from the interference fringes (Fig.~\ref{fig:phase_slip}a); see also~SM for the details of phase slip counting. As shown in Fig.~\ref{fig:phase_slip}b, $N_j$ grows with hold time, yet displaying a peak and a trough at $t\sim 7\,\rm{ms}$ and $\sim 11~$ms, respectively, and afterwards demonstrating an increasing trend. This behavior is indeed responsible for the observed evolution of long-range phase coherence in Fig.~\ref{fig:fig2}d, complementing the dynamical structure formation observed in Fig.~\ref{fig:fig2}c.

In our experiments, finite transverse confinement along the $y$-axis, the initial noise, and the presence of atom loss alter the quench dynamics compared to a pure 1D case (Fig.~\ref{fig:fig1}c). To adequately model the emergent nonequilibrium behavior in our quasi-1D gases, we exploit a 2D GPE model considering these realistic conditions; see SM for details. In our computations, we find prominent density undulations generated from the box edges and traveling towards the center in a sample averaged density, resembling the experimental images (Fig.~\ref{fig:fig2}a) and in line with the theoretical expectation from the corresponding dam-break system~\cite{el2016dam}; see Supplementary Fig.~\ref{fig:figSI6} in~SM for comparison with Fig.~\ref{fig:fig2}. The simulated density-density and two-point phase correlation functions qualitatively reproduce the observed dynamical periodicity and finite-range phase coherence, with the exception that we have not observed the transient improvement of quasi-long-range phase coherence and a clear reduction of phase jumps around $10\sim13~$ms when the DSWs interact with the noise-induced density modulations; see Fig.~\ref{fig:phase_slip}b. At longer times, the time-averaged correlation lengths similarly drop to the healing length scale (see Supplementary Fig.~\ref{fig:figSI6}). 

We now turn to the study of the rephasing dynamics starting from phase-scrambled attractive Bose gases. More specifically, after a sufficiently long hold time $t=50\,\rm{ms}$ when the phase correlation length shrinks to $\sim \xi$ (Fig.~\ref{fig:fig2}d), we perform an interaction quench back, where the scattering length is linearly ramped to its initial repulsive value within a variable ramp time $t_{r}$ (Fig.~\ref{fig:fig1}a). Figure~\ref{fig:fig4}a shows sample images of interference fringes taken at the end of the ramp. Under faster ramps, we observe that the fringes are more frequently interrupted by phase slips, suggesting the proliferation of density defects, that is, dark and gray solitons~\cite{Frantzeskakis_2010}. This can be understood as phase-incoherent domains start to merge at $g>0$ and form density defects at the domain boundaries; see~SM and Supplementary Fig.~\ref{fig:figSI7}. Considerably larger phase-coherent domains are observed with slower ramps. For $t_{r} \gtrsim 132\,\rm{ms}$, the two-point phase correlations (Fig.~\ref{fig:fig4}b) show re-established quasi-long-range coherence. The length scale is remarkably comparable to that measured in the initial quasi-1D gas. This suggests that no significant heating was introduced during the attractive interaction stage, even after an attractive gas has seemingly lost its quasi-long-range phase coherence; see also SM for the analysis of energy per atom following the rephasing dynamics.

In Fig.~\ref{fig:fig4}c, we show the defect counting following the interaction ramp-back. With faster ramps, $N_j$ roughly remains constant. For slow ramp-up times $t_r \gtrsim 10~$ms, the defect number gradually reduces, leading to improved phase coherence observed in Fig.~\ref{fig:fig4}b. With the longest $t_r\sim 1~$s, $N_j$ reduces by around fourfold compared with the defect number before the ramp. 

To qualitatively understand the observed rephasing behavior, we employ GPE simulations. We point out that, in an integrable 1D model, the phase slips generated in the attractive interaction regime convert to stable solitonic excitations regardless of the ramp-up time. Hence, $N_j$ must be conserved. The 2D simulations, on the other hand, exhibit qualitatively similar rephasing dynamics as in the experiment after transitioning into the repulsive regime. The simulated number of phase jumps is depicted in Fig.~\ref{fig:fig4}c, qualitatively reproducing the observed defect statistics except for long $t_r \gtrsim 100~$ms where atom losses primarily caused by three-body recombination become apparent; see SM for the atom loss analysis. 

It is crucial to emphasize the role of the higher dimensionality in the rephasing dynamics. As the interaction increases during the ramp-up, the healing length decreases to several times smaller than the transverse oscillator length $l_y$ of the box trap. This allows solitonic excitations to convert into vortex dipoles~\cite{KIVSHAR2000117,Kevrekidis2015, tamura2023observation}, which can subsequently decay into sound waves via their collisions~\cite{kwon2021sound} and through drifting out from the gas~\cite{kanai2021true}, see also~SM for sample events observed in the 2D GPE simulations. We attribute the observed rephasing in Fig.~\ref{fig:fig4} to these mechanisms. Importantly, the rephasing is not restricted to quasi-1D systems but represents a generic feature of defect dynamics beyond one dimension, which is in line with observations of potential defect annihilation in the 1D-3D crossover regime~\cite{Becker_2013}.

In conclusion, we report the observation of phase-coherent nonlinear dynamics in quasi-1D attractive Bose gases and their rephasing dynamics through interaction quenches between attraction and repulsion. An intriguing extension involves the study of defect formation and annihilation in a wider range of attractive and repulsive scattering lengths, potentially revealing the defect-number scaling in rephasing and domain-growth dynamics. Naturally, further systematic exploration of the role of dimensionality, by controllably modifying the geometry between quasi-1D and progressively more 2D settings, would be of considerable future interest in generating qualitatively different nonlinear dynamics~\cite{dieli2024observation} or in revealing possible connections between the observed rephasing and other spontaneous ordering mechanisms~\cite{kanai2021true}. Additionally, as non-classical density correlations have been observed in attractive Bose gases~\cite{chen2021observation2}, it would be intriguing to explore the role of beyond-mean field effects on the ensuing quench dynamics leveraging more sophisticated numerical methods~\cite{mistakidis2023few}. 

\section*{Acknowledgments}
We thank Qi Zhou and Qiyu Liang for discussions. This work (H.T., S.B., R.L., and C.-L.H) was supported in part by the NSF (Grant \# PHY-2409591), the W. M. Keck Foundation, the DOE QuantISED program through the Fermilab Quantum Consortium, and the AFOSR (FA9550-22-1-0327). This research was also based upon work (P.G.K.) supported by the NSF (Grant \# PHY-2110030, PHY-2408988, and DMS-2204702) and the Simons Foundation (SFI-MPS-SFM-00011048), and was partly conducted while P.G.K. was visiting the Okinawa Institute of Science and Technology through the Theoretical Sciences Visiting Program, and Seoul National University through the Fulbright Program. S.I.M. acknowledges support from the Missouri University of Science and Technology, Department of Physics, Startup fund and the Army research office under Award number W911NF-26-1-A043.

\clearpage

\renewcommand{\figurename}{\textbf{Supplementary Fig.}}
\renewcommand{\thefigure}{\arabic{figure}}
\setcounter{figure}{0}

\clearpage

\title{Supplementary Material for \\
``Direct observation of long-range many-body coherence in quasi-one-dimensional attractive Bose gases"
}

\maketitle

\subsection{Preparation and detection of quasi-1D gases}\mbox{}\\
We begin the experiment by loading a cesium Bose-Einstein condensate (BEC) into two quasi-1D box traps in $400\,\rm{ms}$, followed by a $500\,\rm{ms}$ wait time to reach equilibrium. The initial scattering length is $105a_{0}$. The boxes are formed using $780\,\rm{nm}$ blue-detuned light~\cite{tamura2023observation}. The vertical box confinement is given by a single node of a $3\,\rm{\mu m}$-period optical lattice along the $z$-axis, while the horizontal confinement is produced by a rectangular repulsive potential barrier of $k_\mathrm{B}\times 60\,\rm{nK}$ height, controlled by a digital mirror device (DMD). Here, $k_\mathrm{B}$ is the Boltzmann constant. The measured transverse trap frequencies are $\omega_{z}\approx 2\pi \times2.2\,\rm{kHz}$ and $\omega_{y}\approx 2\pi \times 68\,\rm{Hz}$ along the $z$- and $y$-axes, respectively.

After initial preparation, we quench the scattering length to a negative value at $t=0$ via a magnetic Feshbach resonance. The scattering length is evaluated from the magnetic field using the conversion~\cite{kraemer2006evidence} $a_{s}(B)=(1722+1.52)\left(1-\frac{\Delta B}{B-B_{0}}\right)a_{0}$, where $\Delta B=28.72~\rm{G}$, $B_{0}=-11.60~\rm{G}$~\cite{chen2020observation}, and $B$ represents the magnetic field strength. After the interaction quench, we observe oscillations in the transverse width at a frequency of $\sim 2\omega_{y}$. The oscillation damps down quickly in $3\,\rm{ms}$. After performing the experiments at various hold times $t$, we obtain in-situ or time-of-flight (TOF) density distribution in the horizontal ($x$-$y$) plane by absorption imaging with a resolution of $\approx 0.8\,\rm{\mu m}$. In the rephasing experiments (Fig.~\ref{fig:fig4}), we perform the same imaging procedure after quenching up the attractive interaction back to the initial repulsive value and holding for $15~\rm{ms}$. 

\begin{figure}[h!]
\centering
\includegraphics[width=0.5\textwidth]{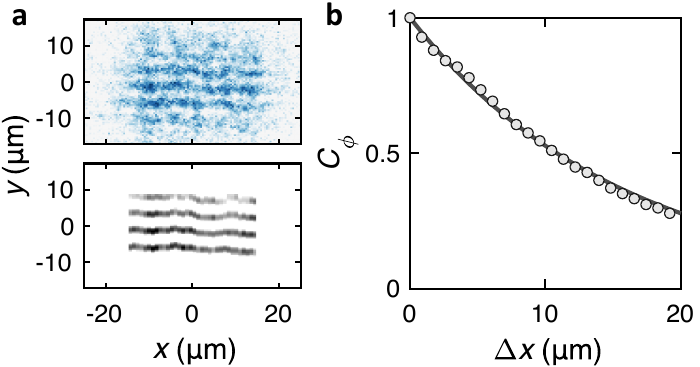}
\caption{\textbf{Experimental interference fringes and two-point relative phase correlation function.} {\textbf{a}}, A single-shot TOF density image (top panel), showing interference fringes of two expanding quasi-1D gases. The central region of the filtered TOF image (bottom panel) is analyzed to extract the relative phase $\phi (x)$ along the longitudinal direction. {\textbf{b}}, Initial two-point relative phase correlation function $C_{\phi}$ at repulsive scattering length $a\approx 105a_{0}$. Solid line is an exponential fit. }
\label{fig:figSI1}
\end{figure}

\begin{figure*}[t!]
\centering
\includegraphics[width=0.95\textwidth]{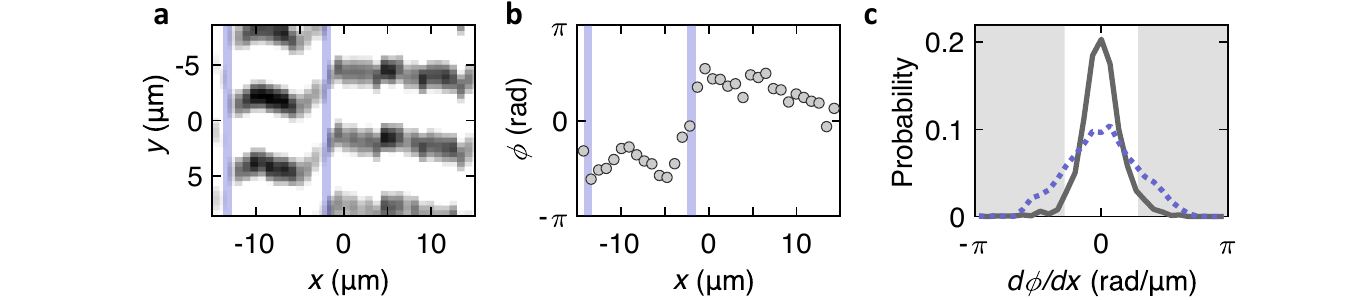}
\caption{
\textbf{Experimental analysis of phase jumps.} {\textbf{a}}, Sample fringe pattern presenting phase jumps observed in the rephasing experiments. {\textbf{b}}, Corresponding relative phase $\phi(x)$. Vertical lines indicate the locations of detected phase jumps. {\textbf{c}}, Event probability of detected phase gradients $d \phi/dx$ recorded before (solid line) and after (dashed line) the $t=30\,\rm{ms}$ hold time at the attractive interaction $-5a_{0}$. The shaded gray area indicates $\left|d \phi/dx \right|>0.29\pi/\rm{\mu m}$.}
\label{fig:figSI2}
\end{figure*}

\subsection{Relative phase measurement}\label{SM:phase}\mbox{}\\
To measure the spatially-resolved relative phase in two parallel quasi-1D gases, we shut off the potential in the $x$-$y$ plane while keeping the vertical confinement, resulting in TOF expansion in the horizontal plane. The quasi-1D gases expand much more rapidly along the $y$ direction, forming interference patterns. During the TOF, we quench the scattering length to nearly zero to avoid nonlinear fringe distortions~\cite{KIVSHAR2000117}. 

Supplementary Fig.~\ref{fig:figSI1}a shows a sample density distribution imaged after $15\, \rm{ms}$ of TOF expansion. To clearly visualize the interference fringes, we apply a bandpass filter along the $y$-axis ($0.5\,{\rm{\mu m^{-1}}}<k_{y}<2\,{\rm{\mu m^{-1}}}$ in the frequency domain). We extract the relative phase $\phi (x)$ from the central part of the fringe patterns ($|x|\leq 15\,\rm{\mu m}$, $|y|\leq8.6\,\rm{\mu m}$) by fitting each density line-cut (along $y$-axis) with a sinusoidal function. 

\subsection{Two-point phase correlation function}\mbox{}\\
We evaluate the relative phase correlation function of two interfering quasi-1D gases as 
\begin{eqnarray}
C_{\phi}(\Delta x) &=& \Braket{e^{i\left[  \phi(x) - \phi (x')\right]}} = e^{-\frac{1}{2} \Braket{ \delta \phi^{2} }}\,, \label{eq:relativephase}
\end{eqnarray}
where $\phi(x) = \varphi_1(x)-\varphi_2(x)$ is the local relative phase between two gases. Also,  $\Braket{\delta \phi^{2}} = \Braket{ \left[\phi (x)- \phi (x')\right]^{2}}$ is the two-point phase variance, $\Delta x=|x-x'|$ refers to the spatial separation, and $\Braket{\cdots}$ denotes ensemble and spatial averaging. We have assumed that  $C_\phi$ is an even function and hence only its real part is non-zero. For degenerate gases with homogeneous density profiles, this is equivalent to measuring the two-point correlation function 
\begin{eqnarray}
C_{\phi} (\Delta x) &=& \frac{\Braket{{\psi}_{1}(x) {\psi}_{2}^{\dagger}(x) {\psi}_{1}^{\dagger}(x') {\psi}_{2}(x')}}{\Braket{\left| {\psi}_{1} (x)\right|^{2}} \Braket{\left| {\psi}_{2} (x')\right|^{2}}}\,,\label{twopoint_correl_phase}
\end{eqnarray}
where ${\psi}_{j}^{(\dagger)}$ is the bosonic field operator and $j\in1,\,2$ designates the gas in each quasi-1D box trap. One can obtain Eq.~\eqref{eq:relativephase} by substituting ${\psi}_{j}\approx \sqrt{n}e^{i\varphi_j(x)}$ into Eq.~(\ref{twopoint_correl_phase}) and using the 1D quasi-condensate density $n$ (assumed to be the same in both boxes) and phase $\varphi_j(x)$. The assumption of equivalent densities is valid for the initial repulsive regime and is a coarse approximation for post-quench gases due to weak density modulation on the background components. For two initial gases with independent phase fluctuations, one can further simplify the correlation function to 
\begin{equation}
C_{\phi}(\Delta x)= |g^{(1)}(x,x')|^{2} \,, 
\end{equation}
where $g^{(1)}(x,x')=\Braket{ \psi (x)\psi^{\dagger} (x')}/n = \Braket{e^{i\left[  \varphi(x) - \varphi (x')\right]}}$ is the first-order coherence function. In this last expression, $C_{\phi}(\Delta x) \in [0,1]$. When it is unity, the gas is fully coherent. Otherwise, coherence losses come into play evidencing the build-up of correlations in the system~\cite{Naraschewski}.

\subsection{Initial correlation length}\mbox{}\\
We measure $C_{\phi}(\Delta x)$ by spatially averaging ${\rm{Re}}[ e^{i \phi (x) -i\phi(x')}]$ over all points within the central region ($|x|,\,|x'|\leq 15\,\rm{\mu m}$). Supplementary Fig.~\ref{fig:figSI1}b displays the recorded initial phase correlation function. We extract the correlation length $L=15.6(2)\,\rm{\mu m}$ using an exponential fit function $e^{-\Delta x / L}$. This corresponds to a coherence length of $L_{{\rm c}}=2L\approx 31~\mu$m in the first-order coherence function.

\begin{figure*}[t!]
\centering
\includegraphics[width=0.95\textwidth]{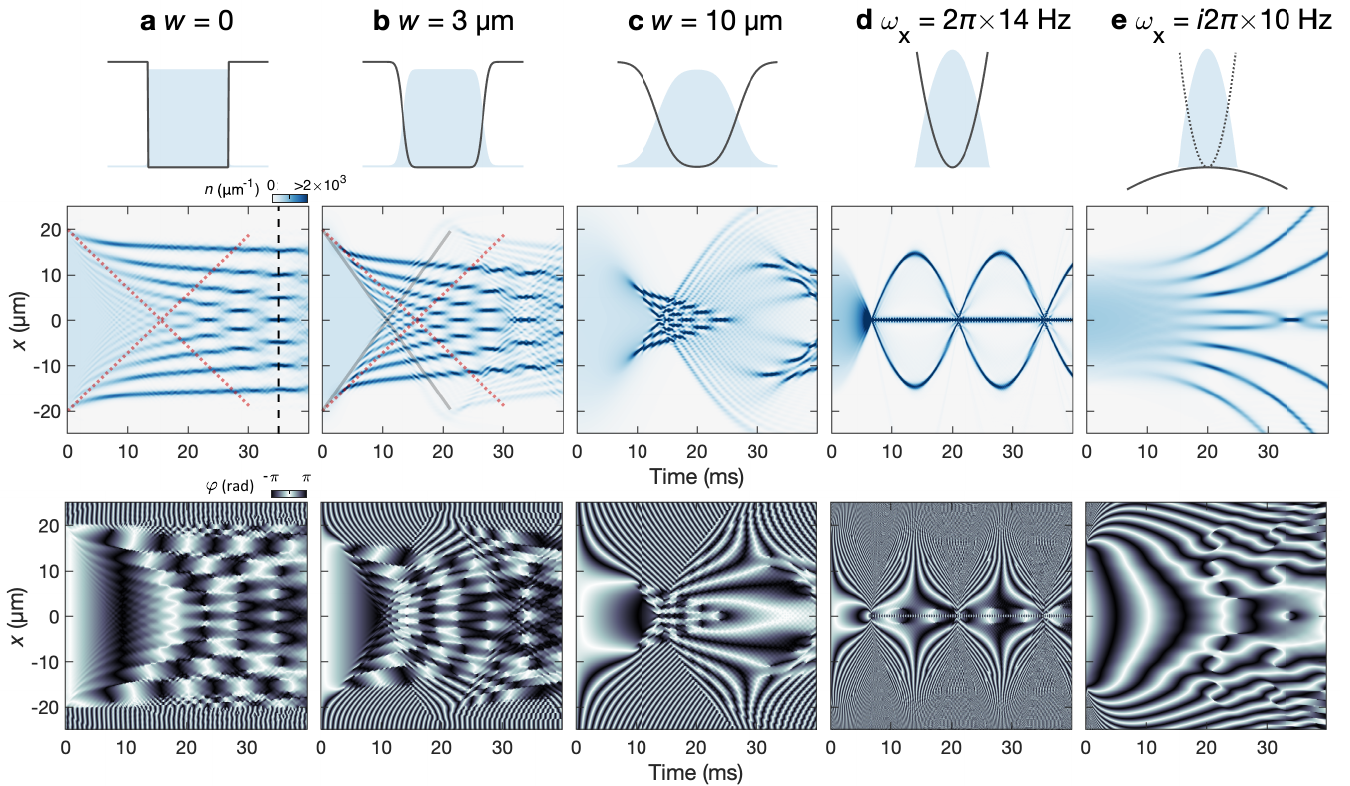}
\caption{\textbf{Numerical 1D simulations of shock wave and soliton generation.} 
{\textbf{a}}, Density (middle) and phase (bottom) evolution following an interaction quench from $a=105a_0$ to $-4.6a_0$ in an infinite hard-wall potential (top) with length $l=40\,\rm{\mu m}$. The vertical dashed line marks $35\,\rm{ms}$, where the density and phase structures are analyzed in Supplementary Fig.~\ref{fig:figSI_1Dcorrelation}. {\textbf{b}} ({\textbf{c}}), Same as {\textbf{a}}, but with widths $w=3\,\rm{\mu m}$ ($w=10\,\rm{\mu m}$), showing faster DSW formation. {\textbf{d}}, Solitary wave formation in a harmonic potential $V(x)=m\omega_{x}x^{2}/2$ with trap frequency $\omega_{x}=2\pi\times 14\,\rm{Hz}$. {\textbf{e}}, Complex solitonic structure formation in an expulsive potential with $\omega_{x}=i2\pi\times 10\,\rm{Hz}$, where a gas is released from a $\omega_{x}=2\pi\times 30\,\rm{Hz}$ harmonic trap upon the interaction quench. The red dotted lines in {\textbf{a}} and {\textbf{b}} show the theoretically predicted flow speed $v$ assuming hard-wall confinement and initial density $400~\rm{\mu m^{-1}}$. The gray solid lines in {\textbf{b}} provide a guide to the eye for monitoring the speed of flow emitted from the wall.}
\label{fig:figSI3}
\end{figure*}

Our sample preparation procedure appears to initiate parallel quasi-1D gases in a \emph{prethermalized state}. For two segmented 1D gases emanating from a zero-temperature BEC, the projected phonons coherently evolve, yet appear dephased in the long-time limit, reaching a prethermalized state~\cite{gring2012relaxation,langen2013local,geiger2014local}. The time-averaged variance of the relative phase is given by $\Braket{ \delta \phi^{2}} = \frac{mg }{\hbar^2}\Delta x$. The effective coherence length $L_{{ \rm pre}}={2\hbar^2}/(mg)\approx 35\,\rm{\mu m}$ matches well to the measured $L_{{\rm c}}\approx 31~\mu$m. On the other hand, if we consider initial phase fluctuations under thermal equilibrium~\cite{stimming2011dephasing}, the expected coherence length in a quasi-condensate is $L_{{\rm th}}=2 \hbar^2n/(mk_{B}T)$. Comparing $L_{{\rm th}}$ with the measured $L_c$, the estimated initial temperature would reach $T\approx 94\,\rm{nK}$, exceeding the energy barrier provided by the confining wall potential.

\subsection{Phase slip detection}\mbox{}\\
From the relative phases $\phi (x)$ obtained from the interference fringes (Supplementary Figs.~\ref{fig:figSI2}a and b), we count the number of phase slips. We first unwrap $\phi (x)$ to remove $2\pi$ phase jumps and evaluate the phase gradient $\frac{d \phi}{d x}$. We then compare $\left|\frac{d \phi}{d x}\right|$ with an empirically selected lower bound threshold of $0.29\pi\,/\rm{\mu m}$ to count phase slips. This threshold refers approximately to the edges of the initial event probability of $\frac{d \phi}{d x}$ (Supplementary Fig.~\ref{fig:figSI2}c) and is chosen to avoid falsely detecting smaller phase distortions as phase jumps. We have applied the same detection criteria to the entire data set presented in Figs. \ref{fig:phase_slip} and \ref{fig:fig4}. The validity of the phase slip detection using 2D GPE simulations is also discussed below. 

\subsection{Dispersive shock waves in 1D GPE simulations}\mbox{}\\
To understand various effects on the dynamics of dam-break flow in a simplified way, we perform time-dependent numerical simulations of the 1D Gross-Pitaevskii equation (GPE):
\begin{eqnarray}
i\hbar \frac{\partial \psi }{\partial t}  &=& \left[ -\frac{\hbar^2}{2m} \frac{\partial^2}{\partial x^2} +g\left| \psi  \right|^2 + V(x) \right]\psi ,
\label{eq:1DGPE}
\end{eqnarray}
where $\psi(x,t)$ represents  the 1D mean-field wavefunction and $V(x)$ is the potential well. 

Dynamics of quenched box-trapped gases is in accordance with the expectation from the underlying dam-break problem. In a situation where the confining potential has infinite hard walls and a length $l$ (Supplementary Fig.~\ref{fig:figSI3}a), DSWs from the boundaries are expected to propagate at a dominant speed $v = \frac{2\hbar}{m\xi}$ (dotted line)~\cite{NLTY2016v29p2798}, irrespective of the box length. This is indeed well captured by the simulated dynamics. Here, $\xi = \hbar/\sqrt{2mn_{0}|g|}$ is the healing length and $n_{0}$ is the initial homogeneous density. The counter-propagating flows weakly interfere with each other, and apparently different structures appear at around $t_{c} = l/(2v)$. We find that Peregrine soliton-like structures spontaneously form initially around $x=0$, while the number of such structures dynamically increases, in line with the theoretical prediction of~\cite{el2016dam}. Those are represented by multiple near $\pi$-phase jumps across each density peak and its tails. The generated density peaks near $x=0$ can be well approximated by the analytical Peregrine waveform. At $35\,\rm{ms}$ (vertical dashed line), well after $t_{c}$, Peregrine soliton-like structures transiently form peaks and weak secondary humps. These are readily visible in the respective single-sample density correlation function $c_{n}=\frac{\Braket{n(x)n(x+\Delta x)}_{x}}{\Braket{n(x)}_{x}\Braket{n(x+\Delta x)}_{x}}-1$ depicted in Supplementary Fig.~\ref{fig:figSI_1Dcorrelation}a. Here, $\Braket{...}_{x}$ denotes spatial averaging. Furthermore, in Supplementary Fig.~\ref{fig:figSI_1Dcorrelation}b, the corresponding single-sample phase correlation function $c_{\phi}=\Braket{e^{i\left[\varphi(x)-\varphi(x+\Delta x) \right]}}_{x}$ exhibits a spatially oscillating structure, reflecting the alternating $\sim \pi$ phase jumps within a single Peregrine soliton.

\begin{figure}[t!]
\centering
\includegraphics[width=0.5\textwidth]{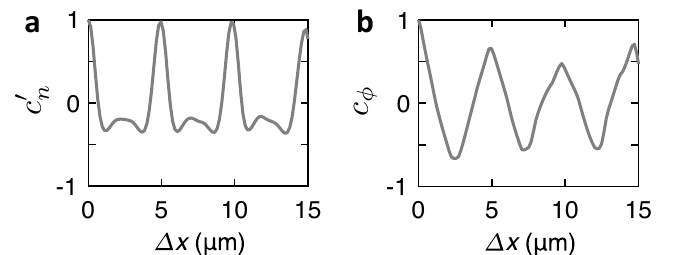}
\caption{\textbf{Simulated two-point correlations.} 
Rescaled density correlation function, {\textbf{a}}, $c_{n}'=c_{n}/c_{n}(0)$ and phase correlation function, {\textbf{b}}, $c_{\phi}$ obtained from the single-shot 1D GPE simulation (Supplementary Fig.~\ref{fig:figSI3}a) at 35 ms.}
\label{fig:figSI_1Dcorrelation}
\end{figure}

In experiments with finite optical resolution, the box wall potential is well approximated by
\begin{eqnarray}
V(x) &=& \frac{U_{0}}{2} \left[ 1 + {\rm{erf}}\left( \frac{\left| x\right| - l/2}{w}\right) \right],\label{soft_poten}
\end{eqnarray}
characterized by width $w$ and dictated by the error function ${\rm{erf}} (x)$. The simulated 1D GPE dynamics reveals that the finite width of the wall potential accelerates the evolution of the counterpropagating DSWs which is evident by comparing Supplementary Figs.~\ref{fig:figSI3} a (hard-wall), b (with $w=3\,\rm{\mu m}$ equivalent to the experimental condition), and c ($w=10\,\rm{\mu m}$ where the potential tends to become harmonic). However, similarly to the hard-wall potential, the emitted DSWs collide at the center and subsequently produce an array of Peregrine soliton-like structures. The multiple Peregrine-like structures start to form at slightly different times as $w$ increases. When $w\sim l/2$, the potential shape around $x=0$ is well approximated by a harmonic trap $V(x)=m\omega_{x}^{2}x^{2}/2$. For a harmonic trap (Supplementary Fig.~\ref{fig:figSI3}d), the interaction quench facilitates a strong focusing effect at the center in the form of a stationary solitary wave and emission of two counter-propagating solitary waves oscillating within the trap. Similar behavior is robustly seen in a broad range of $\omega_{x}$, covering ${4\,\rm{Hz}}\lesssim \omega_{x}/(2\pi)\lesssim {10\,\rm{Hz}}$ used in previous soliton-train experiments~\cite{Strecker2002,nguyen2017formation,robbins,cornish2006formation}. The aforementioned response is dramatically modified upon considering a harmonically trapped gas which is released into an anti-harmonic potential, as theoretically discussed in Ref.~\cite{carr2002dynamics}.  In Supplementary Fig.~\ref{fig:figSI3}e, the numerical simulation shows the generation of complex soliton structures.

\begin{figure}[t!]
\centering
\includegraphics[width=0.5\textwidth]{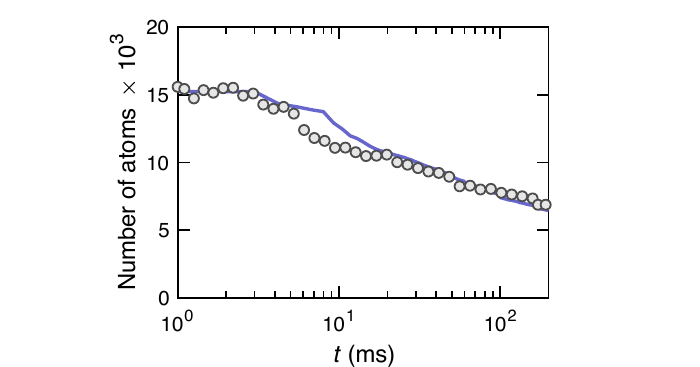}
\caption{
\textbf{Observation of atom loss.} Measured atom number (circles) versus hold time $t$ at the attractive scattering length $a \approx -5a_{0}$. The standard error of the mean is comparable to or smaller than the marker size. The solid line represents the sample-averaged atom number evaluated using the 2D GPE simulations.}
\label{fig:figSI4}
\end{figure}

\begin{figure*}[t!]
\centering
\includegraphics[width=0.95\textwidth]{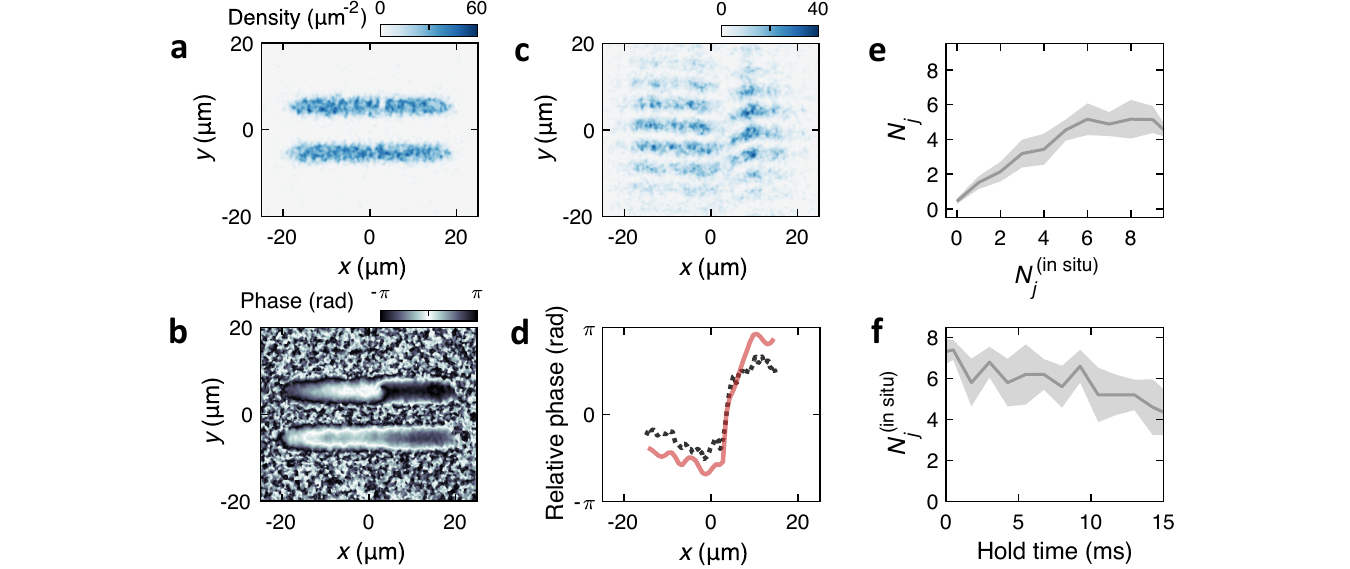}
\caption{\textbf{2D GPE simulations of matter-wave interference.} 
\textbf{a}, An example of simulated density profiles of two parallel quasi-1D samples at the end of the interaction ramp-up procedure depicted in Fig.~\ref{fig:fig1}(a3). \textbf{b}, The corresponding phase profile. \textbf{c}, Simulated density image after $15\,\rm{ms}$ of TOF expansion. \textbf{d}, Relative phase between the two clouds; $\phi(x)$ determined by the interference fringes (red solid line) and $\varphi_{1}-\varphi_{2}$ obtained from the in-situ phase difference (black dotted line), respectively. \textbf{e}, Number of phase jumps, $N_{j}$, identified in the simulated TOF fringes, versus the number of jumps in the in-situ phase profile, $N_{j}^{(\mathrm{in\, situ})}$ (solid line). The shaded area represents the standard error of the mean. Counting statistics is obtained by analyzing in total $\approx 450$ simulation samples following the experimental procedures outlined in Fig.~\ref{fig:phase_slip} or Fig.~\ref{fig:fig4} with various hold or ramp-up times ($t,\,t_{r} \leq 300\,{\rm{ms}}$). \textbf{f}, $N_{j}^{(\mathrm{in\, situ})}$ as a function of the hold time at repulsive scattering length $105a_{0}$ after a $1\,\rm{ms}$ short ramp from $-5a_{0}$ to $105a_{0}$.} \label{fig:figSI5}
\end{figure*}

\begin{figure*}[!t]
\centering
\includegraphics[width=0.95\textwidth]{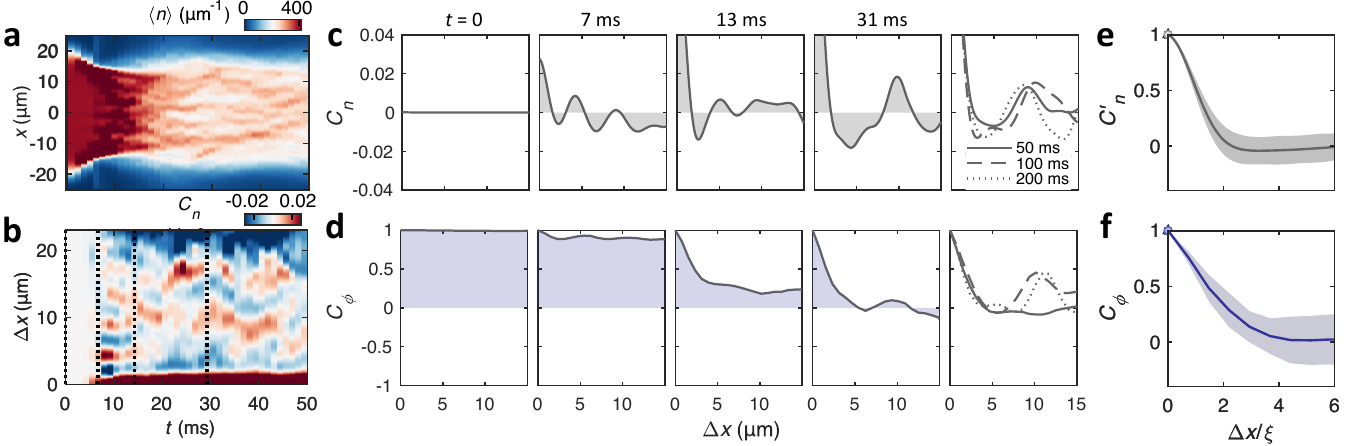}
\caption{\textbf{Dynamical density modulations and phase coherence in 2D GPE simulations.} {\textbf{a}}, Time evolution of the sample-averaged density profile after a quench of the s-wave scattering length from $a = 105a_0$ to $-4.6a_0$. The plotted 1D density along the $x$-axis results from integrating the 2D density over the transverse direction. {\textbf{b}}, Dynamics of density-density correlation function $C_{n} (\Delta x)$ with two-point separation $\Delta x$. {\textbf{c}}, $C_{n} (\Delta x)$ at the characteristic times marked by the dashed lines in {\textbf{b}} and at the indicated long hold times. {\textbf{d}}, Two-point phase correlation function $C_{\phi} (\Delta x)$, evaluated at the same times as {\textbf{c}}, from numerically simulated interference patterns of two independent samples. {\textbf{e}}, $C’_{n}=C_{n}(\Delta x)/C_{n}(0)$ and {\textbf{f}}, $C_{\phi}$ versus rescaled length $\Delta x / \xi$. Both $C'_n$ and $C_{\phi}$ are time-averaged over samples after hold time $t=50\,\rm{ms}$ up to $t=300~$ms. The shaded areas represent the standard deviation from the mean. The simulations follow the experimental procedures and parameters as those used in Fig.~\ref{fig:fig2} and account for the finite resolution effect; see~SM.}
\label{fig:figSI6}
\end{figure*}

\begin{figure*}[t!]
\centering
\includegraphics[width=0.95\textwidth]{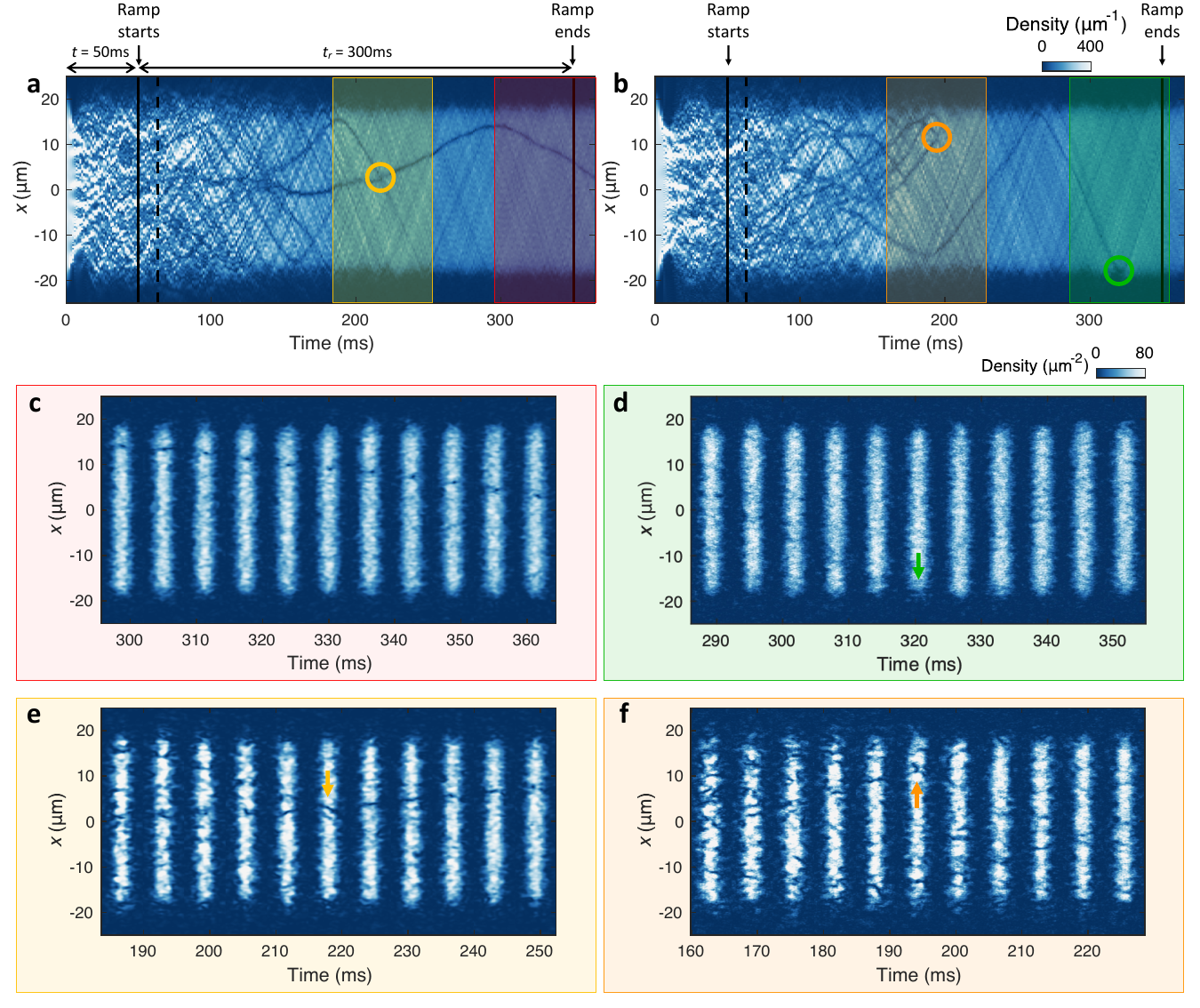}
\caption{\textbf{Defect annihilation and propagation dynamics identified in a quasi-1D gas in 2D GPE simulations.} \textbf{a} and \textbf{b}, Simulated time evolution of density profiles after an interaction quench-down from $a= 105a_{0}$ to $-4.6a_{0}$ and with different initial random noise. During the time evolution, the scattering length is kept constant for a hold time $t=50~\rm{ms}$ and then is ramped back to the initial value in $t_{r}=300~\rm{ms}$, as indicated. Dashed line represents a time at which the scattering length crosses the zero value. The 2D density profiles are integrated over the transverse ($y$-)axis and the resulting 1D profiles are displayed along the $x$-axis. Color shaded areas in \textbf{a} (\textbf{b}) highlight events of sample defect dynamics, where the 2D densities within the corresponding time periods are plotted in \textbf{c} (red) and \textbf{e} (yellow) [\textbf{d} (green) and \textbf{f} (orange)], respectively. \textbf{c}-\textbf{f}, Time evolution of 2D density showing highlighted events: propagation of a vortex dipole-like defect (\textbf{c}), collision of a vortex dipole-like defect with the boundary (\textbf{d}), and inelastic defect collision events (\textbf{e} and \textbf{f}). Arrows indicate the locations of collision events, same as the circles in \textbf{a} and \textbf{b}.}
\label{fig:figSI7}
\end{figure*}
\subsection{2D GPE simulations}\mbox{}\\
The 1D simulations offer an idealized picture of the experimentally observed dynamics due to several reasons. Firstly, in our experimental platform, the confining potential does not completely suppress transverse excitations (along $y$-axis) following the interaction quench. Secondly, there is initial noise (with quantum and thermal origins) in the initial gas. Thirdly, three-body loss is inevitably present in ultracold gases, especially so with higher densities. For these reasons, next we perform  2D GPE computations taking into account the aforementioned contributions in order to more accurately simulate the experimental conditions. The 2D confining potential now reads  $V_\mathrm{2D}(x,y) = V(x)\left[1-{\rm{exp}}\left( -\frac{2y^2}{w_{y}^{2}}\right)\right]$, where $w_{y}$ is the transverse box width and $V(x)$ is described by Eq.~(\ref{soft_poten}). We have added an empirical three-body loss term $-i\frac{\hbar}{2}L_{3}^{\rm{(2D)}}|\psi|^{4} \psi$ to the 2D GPE to phenomenologically capture the experimentally observed slow atom loss. The effective three-body loss coefficient is given by $L_{3}^{\rm{(2D)}}=L_{3}/3! \times 1/(\sqrt{3}\pi l_{z}^{2})$. We have adopted a small 3D recombination loss rate, $L_3=5.5\times 10^{-30}~\rm{cm^{6}/s}$, to match the simulated atom number with the experimentally measured atom loss curve as presented in Supplementary Fig.~\ref{fig:figSI4}. Note that the adopted value of $L_3$ is significantly smaller than those observed in Refs.~\cite{banerjee2024collapse, chen2020observation}, where the gases undergo wave collapse in quasi-2D and suffer significant atom loss. 

To capture the experimental conditions, we also include initial density fluctuations in the pre-quench gases. For this, we empirically seed noise to the GPE predicted zero temperature ground-state density profile $n_{0}=|\psi|^2$. This is achieved through $n = n_{0}\left( 1+ \eta \right)$, where the noise term $\eta = \sum_{k} \frac{n_k}{N_a}\sin \left( kx+\delta_{k} \right)$. Here, $N_a$ is the total particle number, $k$ represents the momentum wavenumber, and $\delta_{k}$ is a random phase uniformly distributed within $[-\pi, \,\pi]$. Furthermore, to involve phase noise we consider $\varphi = \sum_{k} \varphi_k\sin \left( kx+\delta_{k} \right)$. The amplitudes $n_k=\Braket{|n(k)|}$ and $\varphi_k \approx \Braket{|\phi(k)|}/2$ are obtained from the Fourier spectrum of the experimentally measured in-situ density and relative phase profiles, respectively. These amplitudes are further rescaled to match the simulated quench dynamics with experiments. We then evolve the obtained 2D wavefunction $\psi =\sqrt{n}e^{i\varphi}$ for a few milliseconds. Subsequently, this wavefunction is employed as the initial condition for the quench dynamics. Supplementary Fig.~\ref{fig:figSI6} illustrates 2D GPE simulation results in comparison with Fig.~\ref{fig:fig2} (see also the discussion in the main text). We note that the simulated dynamics shows overall qualitative agreement with the experiment (Fig.~\ref{fig:fig2}). However, to match the time scale for developing the characteristic features observed in the density-density correlations, we adopt samples with smaller noise, which results in also better coherence after a long hold time $t\geq 100~$ms. Here, the density-density correlations at $\Delta x \sim 10~\mu$m remain partially phase coherent (the last columns in c and d). The dynamics of phase-slip counts (solid lines in Fig.~\ref{fig:phase_slip}b and Fig.~\ref{fig:fig4}c) are simulated with the same noise amplitude as measured in the experiment, and qualitatively capture the observation, see also the validity of phase slips in Supplementary Fig.~\ref{fig:figSI5}e.

A static disorder in the optical potential can seed modulational instability, thereby accelerating the formation of deterministic density patterns that compete with noise-seeded density modulations and the DSW development. To quantify the impact of disorder, we perform the same simulations as in Supplementary Fig.~\ref{fig:figSI6}, but introduce a controlled spatial variation in the transverse confinement modeled by a position-dependent Gaussian width $w_{y}(x)=w_{y}+\eta_{w}\delta w(x)$. Here, $\delta w(x)$ is a deterministic width disorder evaluated from the measured transverse-width variation of the sample-averaged pre-quench density profile, and $\eta_{w}$ is an adjustable parameter of the disorder amplitude. For sufficiently large variation, the simulations exhibit fixed, reproducible density patterns with a static modulation periodicity, in clear contrast to the dynamically evolving modulations observed in Fig.~\ref{fig:fig2}c. We find that reproducing the observed dynamical density modulations requires a negligibly small potential disorder amplitude, resulting in $\sim 1\%$ variation in the transverse width of the GPE-predicted zero temperature ground-state density profile. This is qualitatively consistent with the experimentally measured variation ($\sim 6\%$) of the averaged density profile prior to any quenches.

\subsection{Validity of phase slips detection}\mbox{}\\
To fully emulate the relative phase measurements discussed in the main text, we evolve two isolated wavefunctions and interfere them via TOF in the 2D GPE calculations. Supplementary Figs.~\ref{fig:figSI5}a and b depict samples of simulated in-situ density and phase profiles emulating the experimental procedure presented in Fig.~\ref{fig:fig1}(a3). Following the TOF expansion, we convolve the simulated density profiles with finite image resolution; see Supplementary Fig.~\ref{fig:figSI5}c. We then analyze the relative phase $\phi$ from the interference patterns, similarly to the experimental procedure (Supplementary Fig.~\ref{fig:figSI2}). Supplementary Fig.~\ref{fig:figSI5}d shows that the results are in qualitative agreement with the phase difference $\varphi_{1}-\varphi_{2}$ directly extracted from the in-situ phase profile weighted-averaged using the mean density. We note that, as the TOF time increases, small-scale fluctuations can be washed out. Therefore, we limit the expansion time to $15\,\rm{ms}$ in our experiments. To confirm the validity of phase slip detection, we compare the number of phase slips in TOF fringes, $N_{j}$, with the one directly obtained from $\varphi_{1}-\varphi_{2}$, denoted as $N_{j}^{(\rm{in\, situ})}$. Supplementary Fig.~\ref{fig:figSI5}e shows the detected $N_{j}$ averaged over samples having $N_{j}^{(\rm{in\, situ})}$ slips. This comparison indicates nearly one-to-one correspondence between in-situ and interference measurements for smaller $N_{j}^{(\rm{in\, situ})}$. This justifies the validity of our observations at $t\lesssim 40~\rm{ms}$ in Fig.~\ref{fig:phase_slip}b and $t_{r}\gtrsim 10 ~\rm{ms}$ in Fig.~\ref{fig:fig4}c. When the density of phase slips increases, each phase slip appears to be less visible, leading to under-counting, see Supplementary Fig.~\ref{fig:figSI5}e.

In Fig.~\ref{fig:fig4}c, phase slips are measured after a $15\,\rm{ms}$ hold time at the final repulsive interaction. For faster ramps with $t_r \lesssim 15\,\rm{ms}$, $N_j$ is systematically smaller than those of Fig.~\ref{fig:phase_slip}b ($t=50\,\rm{ms}$) measured without the fast quench-up. This is attributed to rephasing during the $15\,\rm{ms}$ hold time at the positive interaction. This is supported by Supplementary Figure~\ref{fig:figSI5}f, showing a gradual decay of defect number as the hold time following the quench-up increases.

\begin{figure}[t!]
\centering
\includegraphics[width=0.5\textwidth]{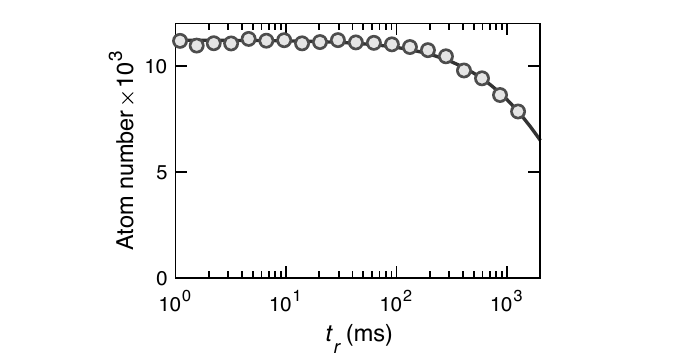}
\caption{
\textbf{Atom number evolution after an interaction quench-up.}
Measured atom number (circles) versus ramp time $t_{r}$ of the scattering length from $-5a_{0}$ to $105a_{0}$. The standard error of the mean is comparable to or smaller than the marker size. The dashed line represents a fit using the analytical model with a fixed three-body coefficient and an adjustable one-body coefficient; see the section `Atomic loss in the interaction quench-up protocol.'}
\label{fig:figSI8}
\end{figure}

\begin{figure*}[t!]
\centering
\includegraphics[width=0.95\textwidth]{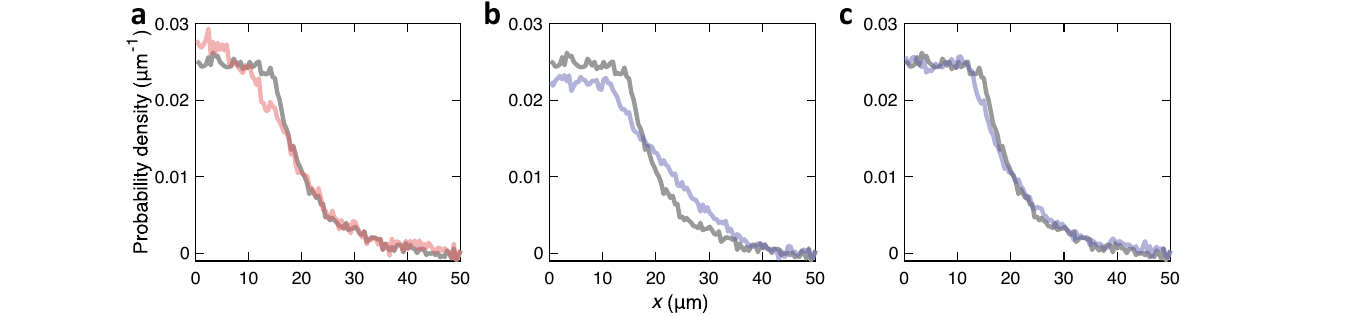}
\caption{
\textbf{Observed density profile after in-plane TOF.}
Longitudinal probability density, namely the sample-averaged TOF density being transversely integrated and normalized by the atom number, recorded before (orange line in \textbf{a}) and after (blue lines in \textbf{b} and \textbf{c}) the interaction quench-up from $-5a_{0}$ to $105a_{0}$ in $t_{r}=1\,\rm{ms}$ and $91\,\rm{ms}$, respectively. In both cases, a fixed hold time $t=50\,\rm{ms}$ is used after the interaction quench-down from repulsion $105a_{0}$ to attraction $-5a_{0}$. In all panels, the gray line represents the probability density prior to the interaction quench-down.
}
\label{fig:figSI_TOFramp}
\end{figure*}

\subsection{Rephasing dynamics}\mbox{}\\
We study the rephasing dynamics by employing 2D GPE simulations closely following the experimental procedure. Prototypical examples of the simulated dynamics are shown in Supplementary Fig.~\ref{fig:figSI7}. In the early times of the ramp, it can be readily seen that the intense density modulation, developed during the epoch of attractive interactions, gradually smoothens out. The coherent domains appear to merge and spontaneously form density defects at the domain boundaries.

At later times of the ramp, as the interaction approaches the final value, these defects noticeably transition into vortex dipoles-like profiles due to the extended dimensionality ($\xi <l_{y}$). For instance, in Supplementary Fig.~\ref{fig:figSI7}a, one can see the propagation of a dark stripe at times $\gtrsim220~\rm{ms}$ propagating with almost un-altered darkness (density contrast). The corresponding 2D images (Supplementary Fig.~\ref{fig:figSI7}c) show two weakly isolated cores, featuring a vortex dipole-like profile. 

These vortical defects are subject to inelastic collisions and annihilation; see, e.g.,~\cite{Becker_2013} for a relevant discussion. During the ramp, when enough defects form, we find intricate collisions between the density defects. At time $\sim218~\rm{ms}$ in Supplementary Fig.~\ref{fig:figSI7}a, two dipole-like defects collide with one becoming indiscernible afterwards; see the 2D images in Supplementary Fig.~\ref{fig:figSI7}e. It is worthwhile to note that in the setting of the latter panel, the structure appears to clearly acquire the form of a single vortex (rather than a vortex dipole) in the confined geometry. Another collision event can be found at $\sim194~\rm{ms}$ in Supplementary Figs.~\ref{fig:figSI7}b and f, showcasing similar dynamical features upon long-time evolution.

The defects can also interact with the boundary and disappear. In Supplementary Fig.~\ref{fig:figSI7}b, a propagating dipole collides with the box boundary at $\sim 320~\rm{ms}$. This dramatically reduces its darkness (contrast), associated with vortex annihilation. The dipole becomes indiscernible after the collision in the corresponding 2D images (Supplementary Fig.~\ref{fig:figSI7}d). Through collisions or interactions with the boundary, the density defects decay into sound, resulting in reduced phase jumps and improved phase coherence. We attribute the observed rephasing to these mechanisms. 

\subsection{Atomic loss in the interaction quench-up protocol}\mbox{}\\
Supplementary Fig.~\ref{fig:figSI8} presents the number of atoms after a quench from attractive to repulsive interactions. It can be seen that the atom number remains nearly constant for $t_{r} \lesssim 100\,\rm{ms}$, within which the phase coherence is well established.

A noticeable loss appears only for longer ramps $t_{r} \gtrsim 100\,\rm{ms}$.
One-body decay due to collisions with background gases ($\lesssim10^{-2}\,\rm{s^{-1}}$) and trap-induced scattering ($\sim 10^{-4}\,\rm{s^{-1}}$) is negligible on the investigated timescales, indicating that three-body recombination is likely the dominant loss mechanism. The associated weak heating may also result in evaporative loss. To place an upper bound on the one-body decay, we consider the rate equation, $\frac{dN}{d \tau}=-L_{1}N-L_{3}^{(\rm{1D})}\int n^{3}\, dx$, accounting for one- and three-body losses under the assumption of homogeneous density profile $n=N/l$. We then fit the data with the solution, $N_{0}e^{-L_{1}\tau}/\sqrt{ 1+2N_{0}^{2}e^{-2L_{1}\tau}L_{3}^{(\rm{1D}) } \tau/l^{2} }$, where $L_{1}$ is the one-body loss coefficient, $L_{3}^{(\rm{1D})}$ is the 1D three-body loss coefficients, $\tau=t_{t} +t_{h}$ is the total time with the fixed interaction hold time $t_{h}=15\,\rm{ms}$ following the interaction ramp, $N_{0}$ is the atom number at $\tau=0$, and $l$ is the box length. Using a fixed three-body coefficient $L_{3}=L_{3}^{(\rm{1D})} \times 3! \times 3\pi^{2}l_{y}^{2}l_{z}^{2}=5.5\times10^{-30}\,\rm{cm^{6}/s}$ (same as the value described in section `2D GPE simulations'), we obtain a one-body lifetime of $1/L_{1}\sim 4 \,{\rm{s}}$, well exceeding the rephasing timescale. This implies that the dominant loss mechanism is three-body recombination. We note that either local density fluctuations or an increase in $L_{3}$ for larger $a_{s}$ may enhance the contribution of three-body recombination.

\begin{figure*}[t!]
\centering
\includegraphics[width=0.9\textwidth]{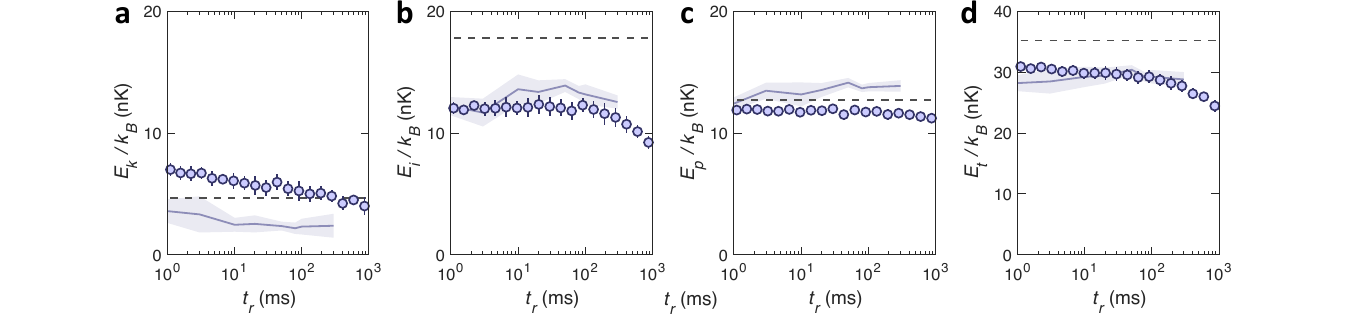}
\caption{
\textbf{Observation of energy per particle in rephased gases.}
Measured energies per atom  (circles) versus ramp time $t_{r}$ from attractive $-5a_{0}$ to repulsive $105a_{0}$ scattering length. \textbf{a}, Kinetic energy $E_{i}$. \textbf{b}, Interaction energy $E_{i}$. \textbf{c}, Potential energy $E_{p}$. \textbf{d}, Total energy $E_{t}=E_{k}+E_{i}+E_{p}$. The dashed lines correspond to the observed initial energies before the interaction quench-down to attraction. The solid lines represent the result of 2D GPE simulations, where the energies are obtained using the same method as for the experimental data. The shaded bands and the error bars indicate the standard error.}
\label{fig:figSI_Energy}
\end{figure*}

\subsection{Analysis of energy per atom}\mbox{}\\
To assess possible heating or cooling during the rephasing process, we extract the involved energies per atom for different $t_{r}$ using in-situ densities and TOF images obtained after ballistic expansion at $a_{s}=0$. 
Supplementary Fig.~\ref{fig:figSI_TOFramp} shows the sample-averaged TOF densities that are transversely integrated and normalized by the atom number. 
The post-ramp distribution (blue line in c) qualitatively matches the initial profile (gray line) prior to the interaction quench, indicating the recovery of the kinetic energy to the pre-quench state. For our quantitative analysis, the kinetic energy is evaluated from
\begin{eqnarray}
E_{k}=\frac{m}{2t_{\rm{TOF}}^{2}N}\left( \Braket{r^{2}} - \Braket{r_{0}^{2}} \right),
\end{eqnarray}
where $\Braket{r^{2}}=\int r^{2} \Braket{n} \,d^{2}r$ is the mean-squared radius after expansion for a duration $t_{\rm{TOF}}$, and $\Braket{r_{0}^{2}}$ is the in-situ value at $t_{\rm{TOF}}=0$. 
From the in-situ densities, we additionally extract the interaction and potential energies,
\begin{eqnarray}
E_{i}=\frac{1}{N}\int \frac{g}{2} \Braket{n^{2}}\,d^{2}r,\,\,E_{p} = \frac{1}{N}\int V_{\rm{2D}} \Braket{n}\,d^{2}r,
\end{eqnarray}
where $N$ is the total atom number and $V_{\rm{2D}}$ is the optical potential (see also section `2D GPE simulations'). Supplementary Figure~\ref{fig:figSI_Energy} summarizes the measured energies per atom.  For ramp times $t_{r}\lesssim 100\,\rm{ms} $, $E_{i}$ and $E_{p}$ remain nearly constant with less than $2\,\rm{nK}$ variation. This trend is inconsistent with evaporative cooling, during which the loss of high-energy atoms simultaneously decreases all energies. In the same time domain, the observed $E_{k}$ gradually returns to the initial value recorded before the interaction quench to attraction, indicating no changes in temperature between those two situations.  The observed recovery of $E_{k}$ complements the reestablishment of long-range phase coherence (Figs.~\ref{fig:fig4}b and c), which occurs without detectable heating, cooling, or dissipation (Supplementary Fig.~\ref{fig:figSI8}). At longer $t_{r}$, the slow atom loss becomes apparent, causing a reduction in $E_{i}$. 
\end{document}